**Title:** Design and Characterization of a Novel Scintillator Array for In Vivo Monitoring During UHDR PBS Proton Therapy

Roman Vasyltsiv[1], Joseph Harms[2], Megan Clark[1], David J. Gladstone[1], Brian W. Pogue[1,3], Rongxiao Zhang[1,4], Petr Bruza[1]

[1]Thayer School of Engineering, Dartmouth College, Hanover, NH
[2]Department of Radiation Oncology, University of Alabama at Birmingham, Birmingham, AL
[3]Department of Medical Physics, University of Wisconsin-Madison, Madison, WI
[4]Department of Radiation Oncology, University of Missouri, Columbia, MO

**Abstract:**
**Background:** Ultra-high dose rate (UHDR) proton therapy has shown promise in normal tissue sparing by enhancing the therapeutic ratio through a method termed the FLASH effect. As in all radiotherapy, accurate in vivo dosimetry is crucial for quality assurance of safe and efficient treatment delivery. However, this remains a challenge for UHDR as existing dosimetry systems lack the spatial and temporal resolution required to verify dose and dose rate in complex anatomical regions, especially for pencil beam scanning (PBS) proton therapy.

**Purpose:** This study aims to develop and evaluate a novel 3D surface dosimetry method for UHDR PBS proton therapy using high-speed imaging of a scintillator array, coupled with stereovision to provide real-time, high-resolution surface dose monitoring during treatment. The spatial, temporal, and dosimetric components of the proposed system are validated via imaging of a custom QA phantom and are compared against treatment planning system (TPS) predictions.

**Methods:** A freely deformable multi-element scintillator array was designed with a single element pitch of 7.5 mm and inter-element gap of 0.5 mm. Scintillation linearity with dose was evaluated along with the variation in scintillator response with increasing imaging and irradiation angles. Water-equivalent depth (WED) testing was conducted to evaluate beam attenuation at two energy levels. Scintillation emission in response to dose delivery was imaged at 1000Hz using a high frame rate camera (BeamSite Ultra, DoseOptics LLC) and the mesh position was monitored via a 2-camera stereovision system. Imaging system setup was validated using a custom 3D QA phantom to assess spatial accuracy and guide systematic setup correction. Stereovision properties of each array element were used to guide angular emission correction, and geometric transformation to beams-eye-view (BEV). Kernel-based residual spot fitting was applied to derive cumulative dose maps which were then compared to the TPS dose profile of a 5x5cm UHDR PBS delivery using 3%/2mm gamma analysis. PBS and maximum dose rate maps were also calculated.

**Results:** System setup achieved an average localization error of 0.62 mm, surpassing the typical 1+ mm threshold used in clinical practice. Intensity correction based on angular information was applied and yielded a cumulative spot dose uncertainty of ~1% (5.428 mGy). The processed dose map was compared to the TPS plan via gamma analysis with 3%/2 mm criteria and showed a 99.9% passing rate, indicating high agreement between the planned and measured dose profiles. The water-equivalent depth (WED) of the scintillator array was measured to be 1.1 mm, minimizing its impact on dose distribution.

**Conclusion:** The novel scintillator array system provides accurate, real-time surface dose monitoring with high spatial and temporal resolution, making it a promising tool for in vivo dosimetry in UHDR proton therapy. Future work will focus on optimizing the system and expanding its application to other modalities, such as photon and electron therapy.

## 1. Introduction

Ultra-high dose rate (UHDR) radiation delivery has demonstrated improved therapeutic ratio via healthy tissue sparing, known as the FLASH effect, both in vitro[1] and in animal studies[2–4]. These promising results have driven the translation of FLASH radiotherapy towards first clinical studies across various modalities, including electron therapy and most notably proton therapy, which is particularly attractive owing its capability of treating deep-seated tumors. Achieving UHDR dose rates with pencil beam scanning (PBS) beam delivery[5] allowed translation to the first FLASH clinical trials[6,7] while maintaining clinical-level control over scanning velocity, dwell times, and instantaneous dose rates. The PBS method, however, introduces a need for novel dose rate metrics[8] that may significantly impact FLASH treatment outcomes due to the complex spatiotemporal modulation of PBS fields. Evaluation techniques, such as local average dose rate (LADR), have been proposed[8] and UHDR quality assurance had to therefore be extended to incorporate these new metrics[9,10]. These requirements impose significant demands on dosimetry systems due to the necessity of achieving both high (sub-millimeter) spatial and temporal (sub-millisecond) resolution concurrently[11]. To this date, only two modalities – strip ion chamber[12] and 2D scintillation imaging[13] have met these requirements in a flat phantom. To the best of our knowledge, there is no system that can perform spatial, time resolved dose and LADR verification *in vivo*. With phase I (safety) PBS proton FLASH clinical trials being completed[6,14] and recent reports of up to 10% dose and LADR dose rate variations[15], the lack of adequate tools for FLASH PT treatment validation remains a major challenge.

The difficulty of performing FLASH PBS delivery validation stems from the scanning dynamics (>20mm/msec scanning speed, <2ms spot dwell time)[15] combined with high current (>500 nA)[16] and small spot size (<5mm FWHM, as a result of increased energy to meet dose rate criteria), introducing temporal and spatial requirements that must be met concurrently. First, the temporal resolution must be faster than 2 ms to resolve the dose rate of a passing-by PBS beam. This can only be achieved with ion chamber, microdiamond, diode, and scintillation type dosimeters[17]; chemical dosimeters including film and alanine dosimeters require *ex post* readout and are incapable of quantifying dose rates. Second, fast beam scanning results in highly spatially heterogeneous dose rate maps especially at surface down to 1 mm[13]. The LADR metric is especially sensitive to alignment with scanning pattern, where dose rate gradients as high as 1100 Gy·s$^{-1}$·mm$^{-1}$ can be observed on the surface[13]. *In vivo*, such alignment accuracy practically disqualifies the use of point dosimeters where sub-mm localization accuracy would be necessary to yield meaningful comparison against a reference (planned) value. The 2D strip ionization chamber array (SICA) has previously been shown to provide 2D time-resolved dosimetry with sub-millisecond temporal resolution[12,18]; however, its 2-mm spacing and reliance on circular symmetry (as the array reports two 1D projections rather than a full 2D view) limits the spatial resolution and accuracy. Furthermore, the flat rigid design of the array makes it unsuitable for *in vivo* use, as it cannot conform to patient anatomy and could present significant surface build-up or beam attenuation effects, which limits the effectiveness in clinical dose and dose rate verification.

Scintillator imaging dosimetry (SID) has been successfully used in UHDR quality assurance (QA)[19–22] and in vivo[23,24], and therefore presents an ideal solution to both applications. Flat scintillator sheet imaging was used for UHDR PBS proton field validation with resolution of 0.5 mm and 1 ms, measuring dose rates up to 500 Gy/s. However, similar to SICA, the rigid scintillator target design did not allow for direct *in vivo* use. Flexible scintillator solutions were proposed[25], but their design lacked methods to fully correct their response with respect to beam and camera optical axis angles as well as address their discontinuous sampling. Single point scintillators were successfully imaged in the conventional clinical setting[23,26], and while these plastic scintillator targets could potentially be used in protons, their rigidity and limited size would lead to aforementioned difficulty with reference dose co-registration. To address these limitations, we developed a UHDR dosimetric solution including a novel discretized scintillating array, which we integrated with a high-frame-rate camera and stereovision imaging. The unique rigid-flex design of the scintillating mesh elements allowed for both sub-millimeter spatial localization in 3D treatment coordinate space, and accurate angular emissivity correction[27]. In this work, we demonstrate the system design and its performance under proton UHDR conditions, highlighting its potential as a reliable tool for in vivo dosimetry in this evolving field.

## 2. Methods:

This section presents a description of the experimental setup, image analysis, calibration data acquisition, and an end-to-end practical demonstration.

### 2.1 Design rationale and overview

An ideal *in vivo* dosimeter shall produce data that can be directly compared to a ground truth reading or the treatment plan. Since treatment planning system (TPS) fields are modeled in a beam's-eye view (BEV), the 2D dose and dose rate profiles captured on the patient anatomy must also be transformed to this same view for proper comparison. This transformation requires determining the 3D distribution of the imaged dose profile and converting it to the TPS frame of reference. In addition, due to angular dependencies introduced by various components of the imaging system, calibration and correction factors are necessary to preserve the intensity-to-dose relationship. For this purpose we implemented a stereovision imaging system, which provides both the geometric transformation of the iCMOS images to BEV, as well as the angular emissivity correction that is crucial to retaining dose linearity. As the mesh deformed and moved with the target surface topology, the cameras mapped both the emitted scintillation intensity as well as the 3D distribution of the array elements and their relative angle with respect to gantry angle and optical axis. By aligning the 3D surface profile with the known orientation of the BEV, the intensity map was transformed to match the TPS frame of reference, and was finally compared to the planned surface dose.

### 2.2 Experimental Setup

The typical experimental setup is shown in Figure 1a, involving the two main components – a flexible scintillator mesh and a remote imaging system. The imaging system consisted of a high frame rate camera and two stereovision modules arranged as depicted in Figure 1b. The high frame rate camera was used to record the scintillation emission, while the stereovision system provided data for 3D mesh reconstruction. The high frame rate camera (BeamSite Ultra, DoseOptics LLC) consisted of an intensified CMOS (iCMOS) fast sensor coupled a with blue-sensitive intensifier, 50mm f/2 lens, and interfaced via 4× CXP-12 link to a PCIe frame grabber, enabling image acquisition at 1000 fps (998 μs frame duration, 2 μs dead time). The high frame rate camera was mounted in the center of a stereovision system which consisted of two FLIR Blackfly-S USB3 modules each equipped with a 50 mm lens, displaced from center by 7.5 cm, and turned inward by 8 degrees to ensure that the target was centered in the image for each camera. The camera was mounted 2 m away from the isocenter at a height of 2.2 m above the treatment couch and focused onto the isocenter.

The scintillator array (US patent US12036421B2) consisted of a mesh of hexagonal elements with a pitch of 7.5 mm and an inter-element gap of 0.5 mm. Each element had a thickness of 1 mm and consisted of a polylactic acid (PLA) supporting base, and a scintillating insert (Penn-Jersey X-Ray, Blue-800 BaFBr Scintillator). Optically opaque PLA walls were created around the perimeter of each element to prevent optical crosstalk between individual elements.

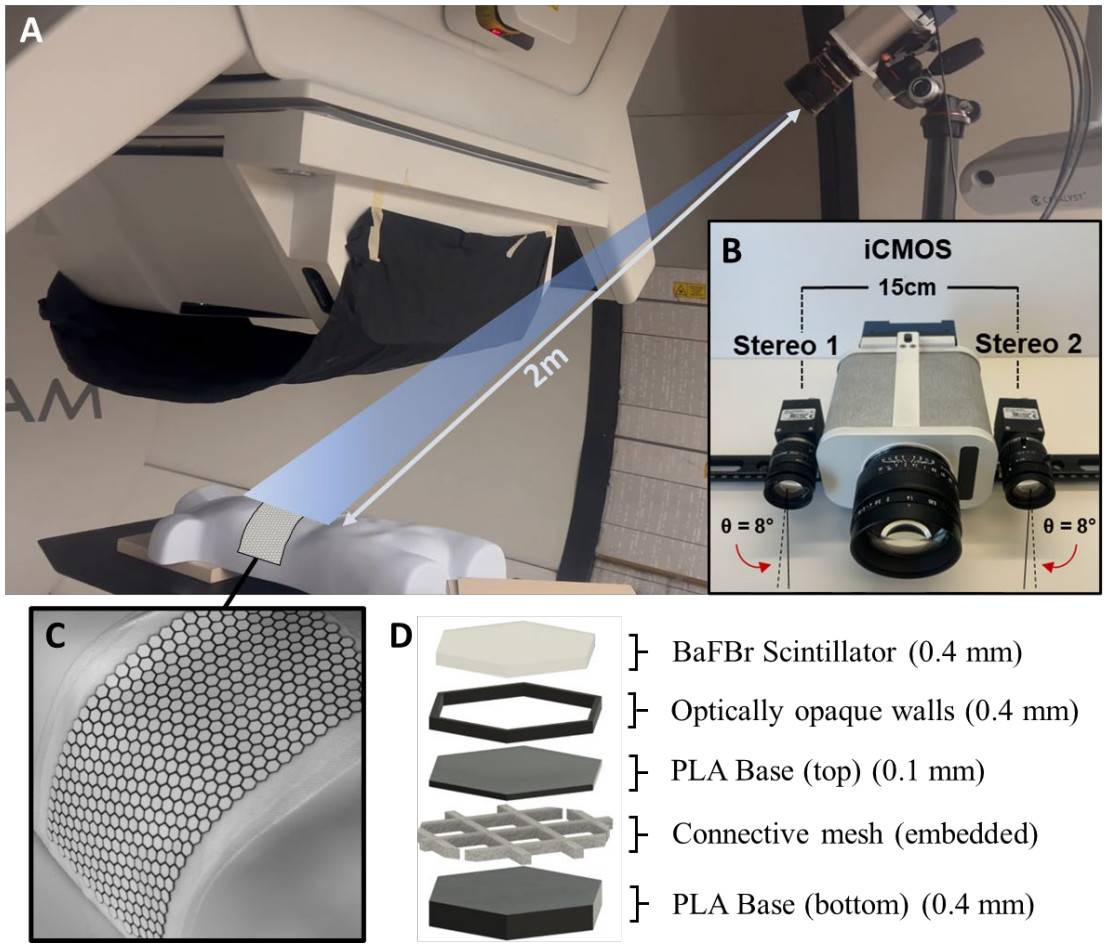

**Figure 1:** (a) Experimental setup showing the novel scintillator array over the patient treatment area, proton gantry delivering a PBS treatment, and an imaging system located at 2 m distance from target. Photo of the imaging system (b) consisting of high frame rate camera and two neighboring stereovision modules. Detailed photo of the scintillator array used for this study is shown (c) along with a breakdown of the element composition (d).

### 2.3 Image Pre-Processing

Every image generated by both camera systems underwent pre-processing prior to any image analysis which was performed in the same way for both calibration and dose imaging stages. Images acquired by the high frame rate camera were saved in an 8-bit format. Each frame collected during the delivery was subject to a dark field and background subtraction to isolate the scintillation intensity map. Images collected by the stereovision modules were also subject to a similar pre-processing procedure where the 8-bit data stream was saved, and a dark field was subtracted from each frame.

### 2.4 Stereovision Camera Calibration

Prior to stereo image reconstruction, the stereovision cameras were calibrated spatially to determine the intrinsic and extrinsic parameters of the imaging setup and account for inter-camera geometry, lens distortions, and the relation of the cameras with respect to the isocenter. A 4x11 point grid located around the gantry isocenter was imaged simultaneously by both cameras; intrinsic calibration dataset included 50 images of the point grid at various angles and locations within the camera's field of view, while extrinsic calibration included a single view of the grid aligned with the isocenter. MATLAB stereoCameraCalibrator application was then used to load and rectify the image pairs by matching the centroids of the point grid in each image pair. The metric of pixel error was used to evaluate calibration results and was set to 0.25 pixels before the intrinsic and extrinsic parameter files were extracted. The position and orientation of the stereo modules was kept consistent for all of the experiments.

### 2.5 Image Fusion and Geometric Correction of Scintillation Images

As described in Section 2.1, the scintillation images from high frame rate camera had to be geometrically transformed to the beam's eye view for dosimetric comparison against the plan. This process is outlined in Figure 2. During the stereovision reconstruction procedure, the scintillating elements were first segmented within each stereo frame using local contrast enhancement and adaptive thresholding. The stereo images from both cameras were then rectified using the known intrinsic and extrinsic camera parameters and placed in the same coordinate space. The centroids from the corresponding elements between the two images were determined and their separation was quantified to create a disparity map. The distance value of the separation was considered together with the corresponding coordinates in image space ($m$, $n$) to create a point cloud (PC) of their position in 3D space ($x$, $y$, $z$). The transformation matrix obtained here was then used to map the scintillation images from the high frame rate camera to the newly defined BEV by matching the centroids of the individual elements in the scintillation image to their analogs in either the left or the right stereovision images. This transformation mapped points in the scintillation image to an image map orthogonal to the gantry output central axis. Local weighted mean registration of the scintillation image yielded a new beam's eye view image of an intensity $I(x', y', t)$, where $x'$ and $y'$ are coordinates in the BEV plane and $t$ is discrete time interval of each high-speed camera frame.

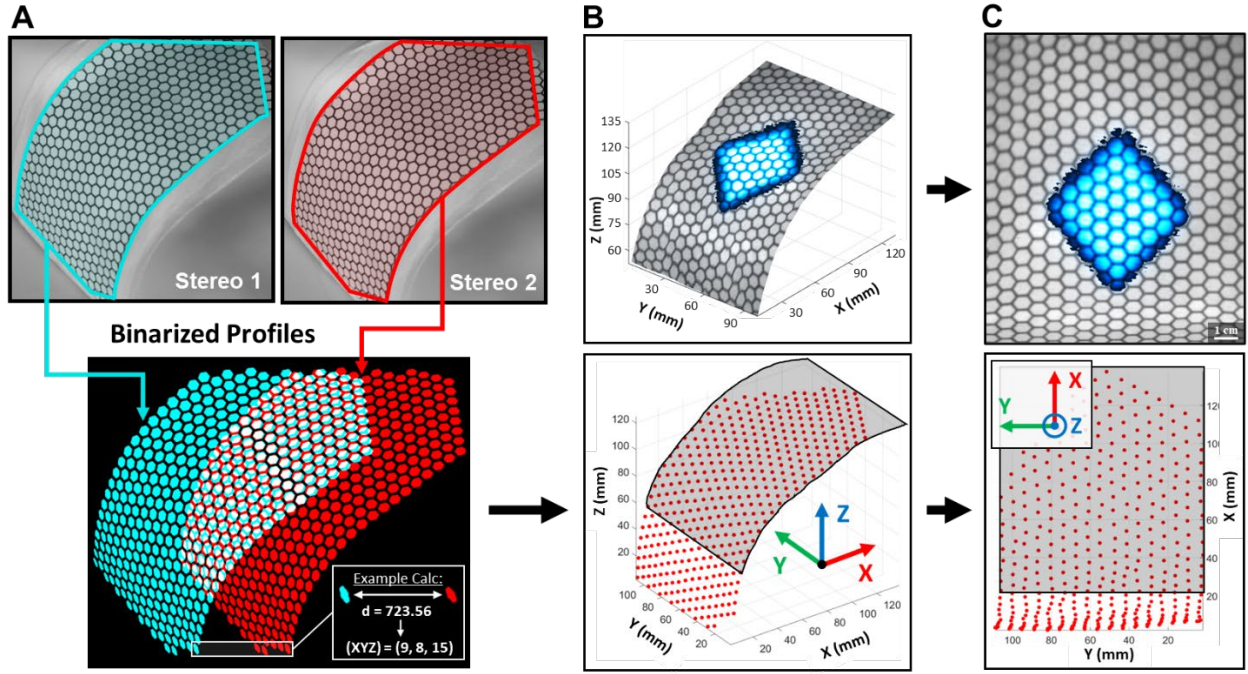

**Figure 2:** Image correction process is shown, using the (a) binarized profiles of the stereovision images to define a (b) 3D centroid point cloud. By translating the point cloud orientation to BEV, the centroid markers are used as anchors to (c) deformably register the scintillation image into BEV via local weighted mean registration.

### 2.6 Scintillation Intensity Correction and Interpolation

The emission intensity of scintillating sheets has been shown to vary based on the viewing angle and irradiation angle, thereby needing correction factors to regain their relationship to dose[27]. Imaging correction factor $k_\theta$ and gantry correction factor $k_\varphi$ were therefore used in this work to retain this relationship, described further in Section 2.7.1. Scale factor *d* was further used to translate the image from units of digital number (DN) to Gray (Gy). Correction for the signal loss between elements was addressed by fitting the corrected spot data to a pre-defined single spot kernel, as described later in Section 2.7.2.

### 2.7.1 Angular Correction

The scintillation output was primarily influenced by the angle of viewing and angle of delivery which were addressed by imaging correction factor $k_\theta$ and gantry correction factor $k_\varphi$, respectively. Each spot ($Q_s$) that contributed to the delivery resulted in scintillation emission from several ($H_s$) neighboring scintillating elements. Each of the elements contributing to a single spot was isolated into a masked image that contained only the signal from the element itself $I(m, n)_j$, and their cumulative contribution to that spot was corrected as follows:

$$Q(m,n)_s = \sum_{j=1}^{H_s} d\, k_{\theta_j}\, k_{\phi_j} \cdot I(m,n)_j$$

Where a single spot ($Q_s$) is the sum of $H_s$ number of individual hexagonal elements emitting signal $I(m, n)_j$, and scaled by the intensity to dose correction factor $d$, angular correction factor with respect to view angle $k_\theta$, and angular correction factor with respect to gantry position $k_\varphi$.

To apply the angular correction to each individual scintillating element, we took advantage of their internal rigidity along with the 3D information obtained from stereovision imaging. Shown in Figure 3b, the centroid point cloud was combined with the geometrically corrected scintillation image and stereovision defined segments (masks) to create a segmented element surface mesh. A normal vector was then found for each centroid (shown in blue) by interpolating a mesh surface between the centroid positions and calculating the normal vector at each position with respect to that surface. Working in the treatment coordinate system allowed us then to define directional vectors at angles θ and φ between each centroid normal vector to the gantry (red) and the camera (green), respectively. Each scintillating element was then assigned two correction factors $k_\theta$ and $k_\varphi$ based on empirical lookup tables. The process was repeated for each scintillating element that contributed to each spot in the delivery to derive a map of the corresponding correction factors, as shown in Figure 3c. These scaling factors were then applied to each element post geometric correction.

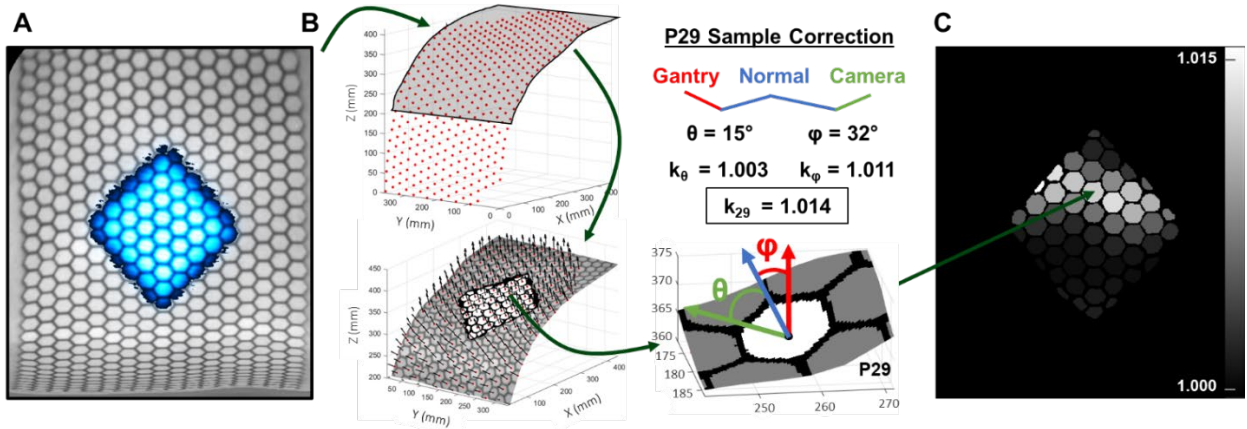

**Figure 3:** Method for angular correction is shown and applied to the (a) geometrically corrected scintillation image. (b) Using 3D information from stereovision surface difference between normal, gantry, and camera angles are determined and related to an intensity correction factor. The process is repeated to all elements containing scintillation signal to create a (c) correction factor image. Data for an arbitrary scintillating element 29 is shown as an example.

### 2.7.2 Discontinuous Sampling Correction

Physical segmentation of the scintillator mesh resulted in signal loss between neighboring elements (Figure 4a), which we addressed by fitting the angle-corrected scintillation profile of each spot to a reference dose kernel (Figure 4b). The kernel (inset in Figure 4b) was acquired by delivering a single 250 MeV PBS spot onto gafchromic film (EBT XD, Ashland) and converting from optical density to dose. The resulting profiles of each spot in the delivery were added together to create the cumulative dose profile as follows:

$$D(n, m) = \sum_{s=1}^{T} R(n + a'_s, m + b'_s) \cdot c'_s$$

Where $R$ is a kernel shifted by parameters $a'$ and $b'$, and scaled by $c'$, optimized to best fit the scintillation data. This was repeated for a total of $T$ spots in the delivery. The $a'$, $b'$, and $c'$ values were obtained by minimizing the unweighted mean of residuals between the fitting kernel R and each respective scintillation spot $Q_s$ at points defined by coordinates of the local maxima $(p_x, p_y)$ in spot $Q_s$:

$$(a_0, b_0, c_0) \rightarrow \min\left(\frac{1}{Z} \cdot \sum_{i=1}^{Z} Q\left(p_{x_i}, p_{y_i}\right)_s - R(p_{x_i} + a, p_{y_i} + b) \cdot c \right) \rightarrow (a', b', c')$$

The fitting algorithm required the definition of initial parameters $(a_0, b_0, c_0)$. Values $a_0$ and $b_0$ were set based on the row and column index of the raw spot signal centroid, respectively, and $c_0$ was set as the maximum value in the single spot scintillation image. This process was repeated for each imaging frame, and the cumulative interpolated image is shown in Figure 4c. The mean of residuals was used to evaluate the goodness of fit for each spot.

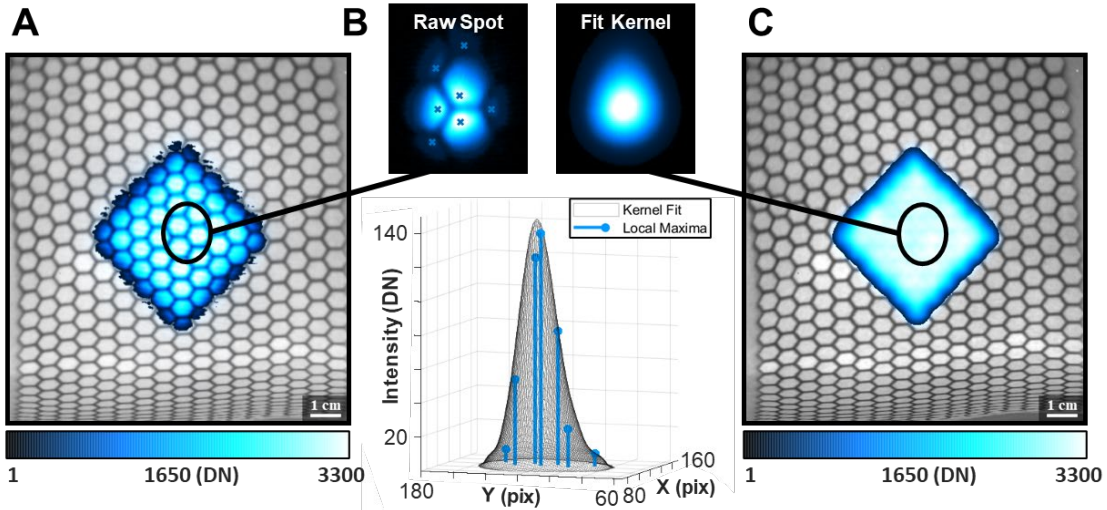

**Figure 4:** (a) Cumulative scintillation image post geometric and angular correction is shown undergoing single spot intensity interpolation. (b) An overlay of the kernel fit over the raw data for an arbitrary scintillation spot 219 is shown, with its location highlighted in the (c) fully interpolated scintillation image.

**2.8 Stereovision calibration and QA**

Because the stereovision information is fundamental to several aspects of the proposed technique, a novel 3D-printed QA phantom was designed to rapidly and thoroughly evaluate the setup and calibration. The stereovision QA phantom shown later in Figure 5a consists of 53 points at varying positions in 3D space, four being placed at the corners at $z = 0$ cm, and the remaining 49 making up a 7x7 grid with 22 mm $x$ and $y$ separation, and with a $z$ separation defined by sampling the following function:

$$z(x, y) = \sqrt{x \cdot y}$$

The viewing elements are white discs with diameter of 10 mm. This setup offers a varied distribution of points in all 3 axes, testing the accuracy of the stereovision reconstruction in the POV of the camera as well as in depth. Validation of point identification was done by finding the Euclidean distance error between the ground truth points and those identified through stereovision imaging. A detailed analysis of distance error was also performed in each of the three principal axes to identify any dominant error trends which could be used to advise any necessary adjustments to the camera setup or location.

**2.9 Correction Factor Acquisition**

Angular correction factors $k_\theta$ and $k_\varphi$ were empirically derived from the dependence of emission intensity on angular deviation. Camera angular dependence was characterized by repeatedly delivering a uniform field to the scintillator array and adjusting the camera incidence angle while maintaining the same array-to-gantry orientation. Intensity readings were taken as an average of 10 scintillator elements every 5 degrees in a range from 0 to 75 degrees, where 0 is normal to the scintillator surface. The emission profile was normalized to 0 degrees to get the relative intensity deviation. The impact of gantry angle deviation was characterized by orienting the scintillator array along a half-cylindrical water-equivalent surface with known angular orientation at each element. The phantom and array were placed at the isocenter and irradiated with a uniform field. The angular orientation of each element was also verified via stereovision imaging as described in Section 2.7.1. To ensure that the intensity variation was only due to the angle with respect to the gantry and not the camera, the data was corrected for the camera angular dependence before the gantry dependence was determined. The emission profile was sampled from 0 to 75 degrees, where the gantry at 0 degrees was normal to the scintillator surface.

The linearity of scintillation intensity with respect to dose was then characterized to determine the dose scale factor $d$. A series of uniform 250 MeV 5x5 cm PBS proton fields were delivered onto a flat scintillator placed on top of film over 5cm of solid water. 5 fields ranging from 2 to 50 Gy were delivered to obtain sufficient data for a linear fit. The average observed scintillation intensity was related to the recorded film dose.

**2.10 Scintillator Array Application**

To characterize the impact of the scintillator array on the beam delivery, the water equivalent depth (WED) of the array was evaluated using a multilayer ionization chamber detector (Giraffe, IBA) at 160MeV and 220MeV. The scintillator array was placed over the detector entrance as a single spot beam was delivered. For both energies, the Bragg peak position was recorded with the scintillator array in place, and the test was repeated with the array removed from the entrance window. The change in Bragg peak deposition depth was used to determine the additional WED presented by the array.

To present a complete application, a pencil beam scanning proton delivery of a 5x5 cm 250MeV, 99nA plan was delivered onto a phantom shown in Figure 1c. A scintillator array was laid over the rib region demonstrating its ability to conform to a target with non-uniform local geometry. The procedure described in sections 2.5-2.7 was applied to obtain the time resolved dose profile. As a form of validation, the final cumulative profile derived from scintillation imaging was compared to a gafchromic film reading of the same field delivered onto a flat surface using gamma analysis. A 3%/2mm threshold was used as defined by standard clinical practice (TG-218). Dose information derived from single frame scintillation signal was also used along with the known constant frame rate to determine the maximum and PBS dose rate profiles[8].

## 3. Results:
### 3.1 Scintillation Spot Fitting Validation

The single kernel fitting used to address the discontinuous sampling between individual scintillator elements was validated via the mean of residuals (MOR) for each spot. A MOR histogram from the full fit stack is shown in Figure 5 indicating high agreement for nearly every imaging frame. A t-location-scale function was fit over the MOR data since the structure of the distribution presented heavier tails compared to a normal distribution, indicating a higher quantity of outliers. The positive tail was mainly attributed to the underfitting of motion frames between dwell positions since these spots naturally further deviated from the assumed form of the ideal kernel. The lower tail was observed as a result of overfitting of partial imaging frames such as the starting and ending frames for each imaged dwell position or the edge frames of each spot delivery, both causing similar deviation from the kernel profile. Despite such deviations from the model, the distribution was found to center about -0.167 mGy with a FWHM of 5.428 mGy which indicates a low predictive error compared to the planned single spot maximum of ~600 mGy.

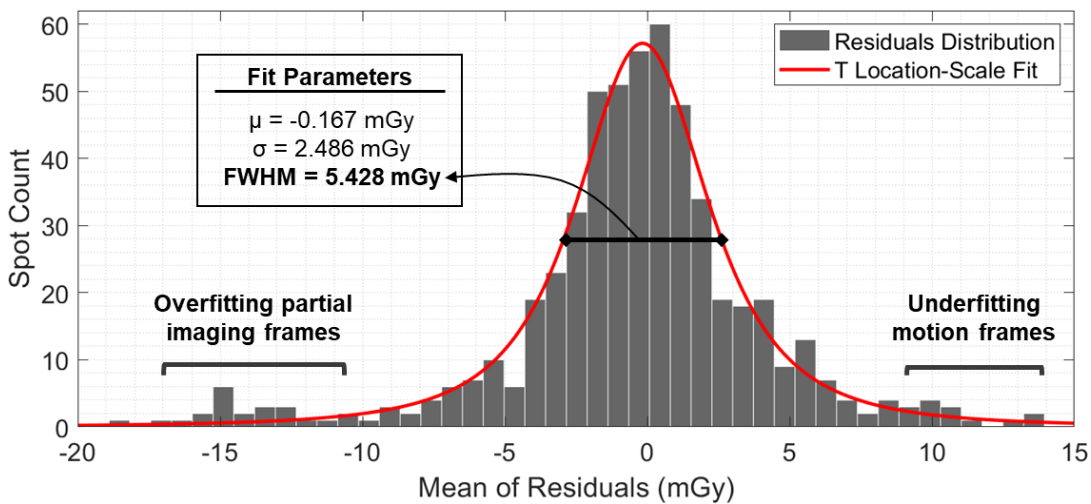

**Figure 5:** Histogram showing the cumulative distribution of mean-of-residual values for every spot in the 5x5 delivery. A t-location-scale function was used to fit the data (FWHM of 5.428 mGy), highlighting the enhanced tails due to overfitting of partial imaging frames and underfitting of motion frames.

### 3.2 Stereovision QA

Following the installation of the imaging setup, the stereovision system was validated as outlined in section 2.8. Expected centroid positions of the phantom were used as reference for the reconstructed positions from the stereo imaging (Figure 6c). Average point deviation from expected position was found to be 0.62 (0.34) mm (Figure 6d), lower than the threshold of 1 mm chosen based on the more strict QA procedures of currently available surface guided systems[28]. Deviations in each of the three axes (Figure 6e) showed no systematic offset, with average single axis errors of -9.1E-14 mm, 9.9E-14 mm, and -7.0E-14 mm for X, Y, Z, respectively.

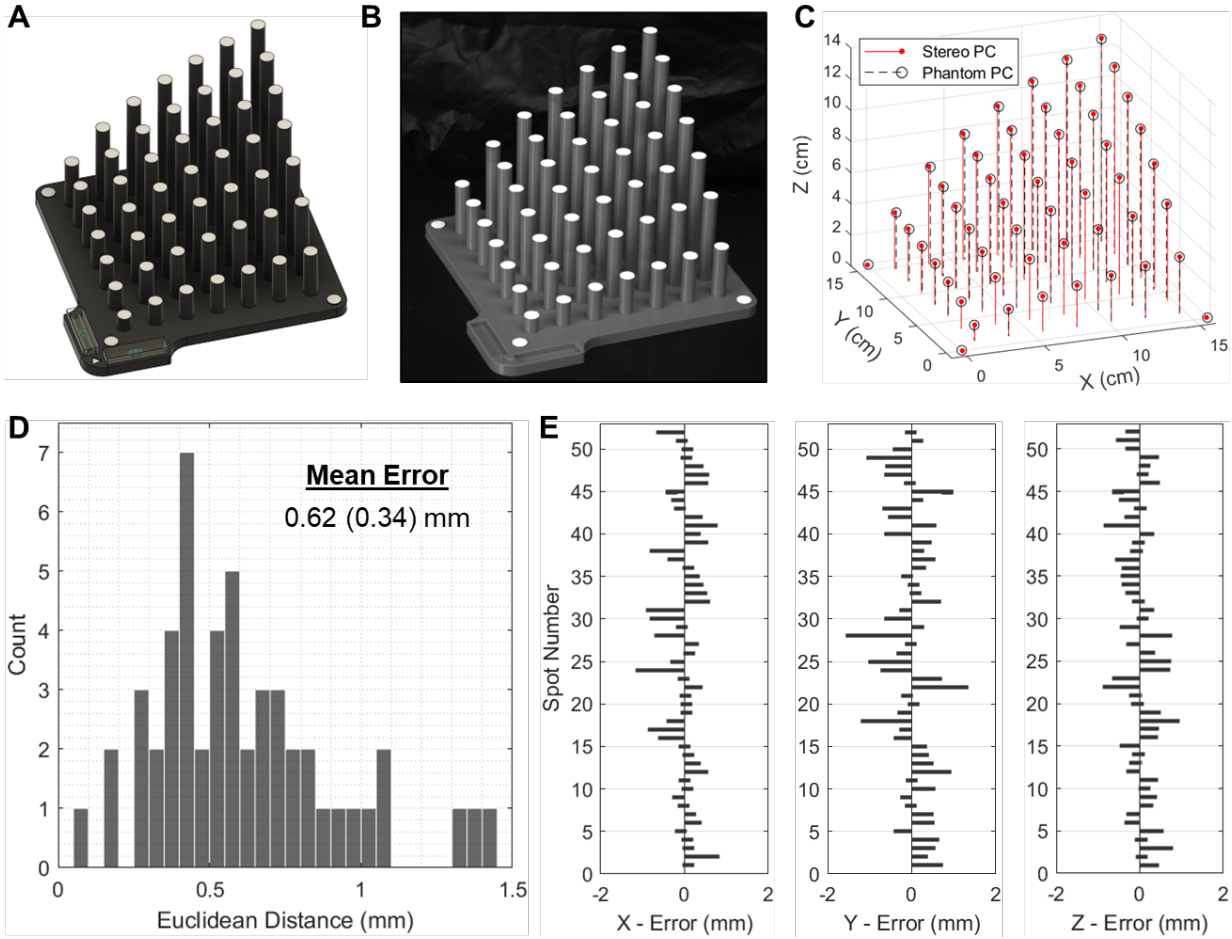

**Figure 6:** Geometric correction procedure showing the system verification using the (a) QA stereovision phantom, (b) the corresponding image from a single stereo camera and (c) point cloud comparison of ground truth phantom locations to the stereovision reconstruction. Centroid localization error was evaluated showing the (d) Euclidean distance error histogram and the (e) single axis distributions.

### 3.3 Angular Correction and Linearity

The dependence of scintillation output on observation angle to the camera and irradiation angle to the gantry was evaluated. Observation angle impact on normalized intensity is shown in Figure 7a and presents no appreciable deviation until ~40 degrees, thereby exhibiting a pseudo-Lambertian profile where the luminance remains constant regardless of observation angle until a certain threshold. No appreciable intensity influence was observed with variation in gantry angle within the tested range (Figure 7b), as all sampled points included 1 within their uncertainty bounds. Scintillation output intensity was characterized with known variation in dose. Figure 7c shows the intensity to dose relationship to be highly linear ($R^2 = 0.9983$) with the y-intercept not being forced to 0, but still including 0 in the uncertainty bounds. The slope from this derived relationship was used to inversely scale the final corrected and fit scintillation map into units of Gy.

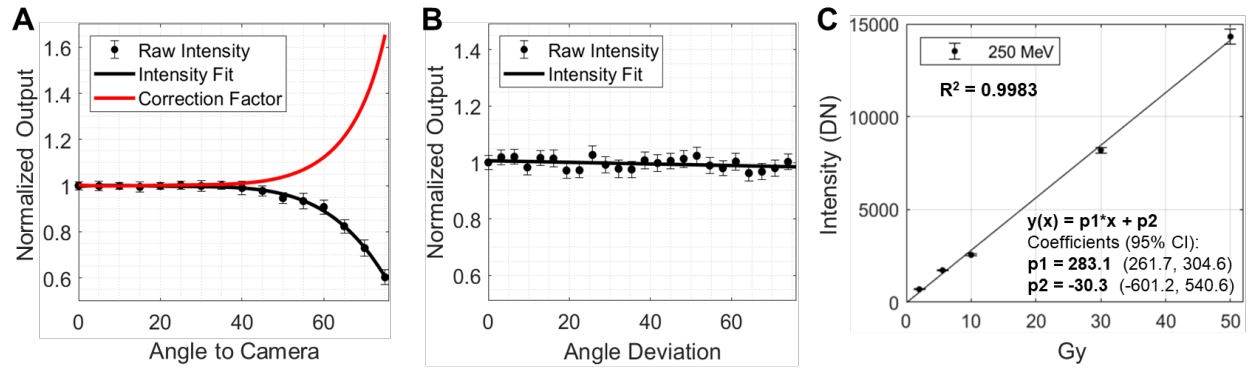

**Figure 7:** Variation in scintillation intensity based on (a) observation angle and (b) irradiation angle was characterized and used to derive angular correction factors for the angular correction process outlined in section 2.7. The dependence of scintillation output on (c) dose was characterized and fit to a linear function showing high accuracy ($R^2 = 0.9983$).

### 3.4 Impact of Scintillator Array on Beam Delivery

Water-equivalent depth testing yielded a WED of 1.1mm for 160 MeV and 200 MeV. For cases where the additional WED would be considered to significantly impact the clinical prescription, the 1.1 mm WED of the scintillator mesh can be accounted for either by simulating patients with the mesh in place or adding it virtually in the treatment planning system like a bolus.

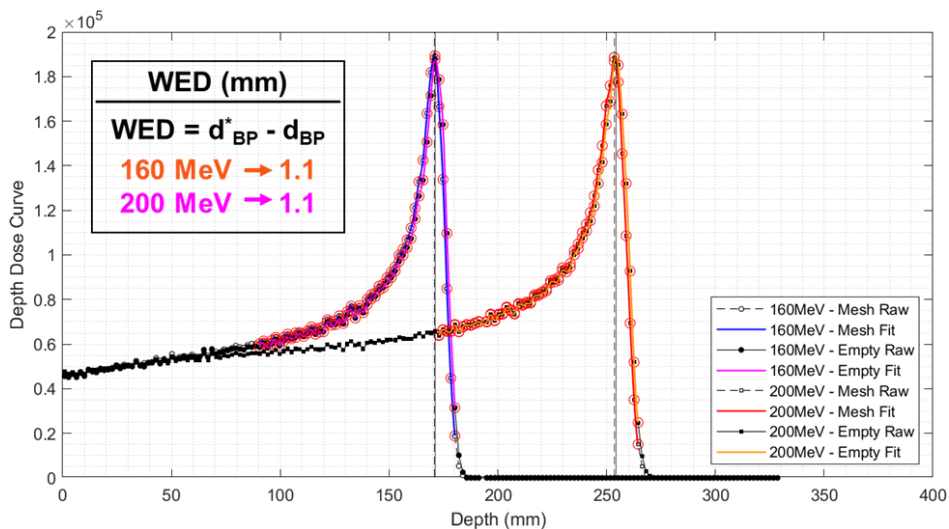

**Figure 8:** Modelled setup used to characterize the water equivalent depth (WED) of the scintillator array by delivering beams of known energy to a multilayer ionization chamber detector. Tests at 160 MeV and 200 MeV similarly indicate a WED of 1.1mm.

### 3.5 Complete UHDR PBS Application

The application of the proposed method is shown in Figure 9 where a uniform 5x5 cm PBS proton field is delivered onto a rib region of an anthropomorphic phantom. Using the geometric information derived from stereovision imaging, the scintillator array point cloud served as a guide to create a surface mesh and overlay the pseudo-colored cumulative scintillation delivery seen in Figure 9a as it would be viewed in the coordinate space of the treatment environment. Following the correction steps outlined in sections 2.5-2.7, the perspective of the delivery was adjusted to beams-eye-view and the signal loss between elements was interpolated to yield a cumulative dose map shown in Figure 9b. The gamma analysis (Figure 9d) done using a 3%/2mm threshold showed a passing rate of 99.9% indicating high agreement between the scintillation and film (Figure 9c) dose maps. The known frame rate of the scintillation imaging approach was also leveraged to derive the maximum and PBS dose rate maps as shown in Figures 9e and 9f.

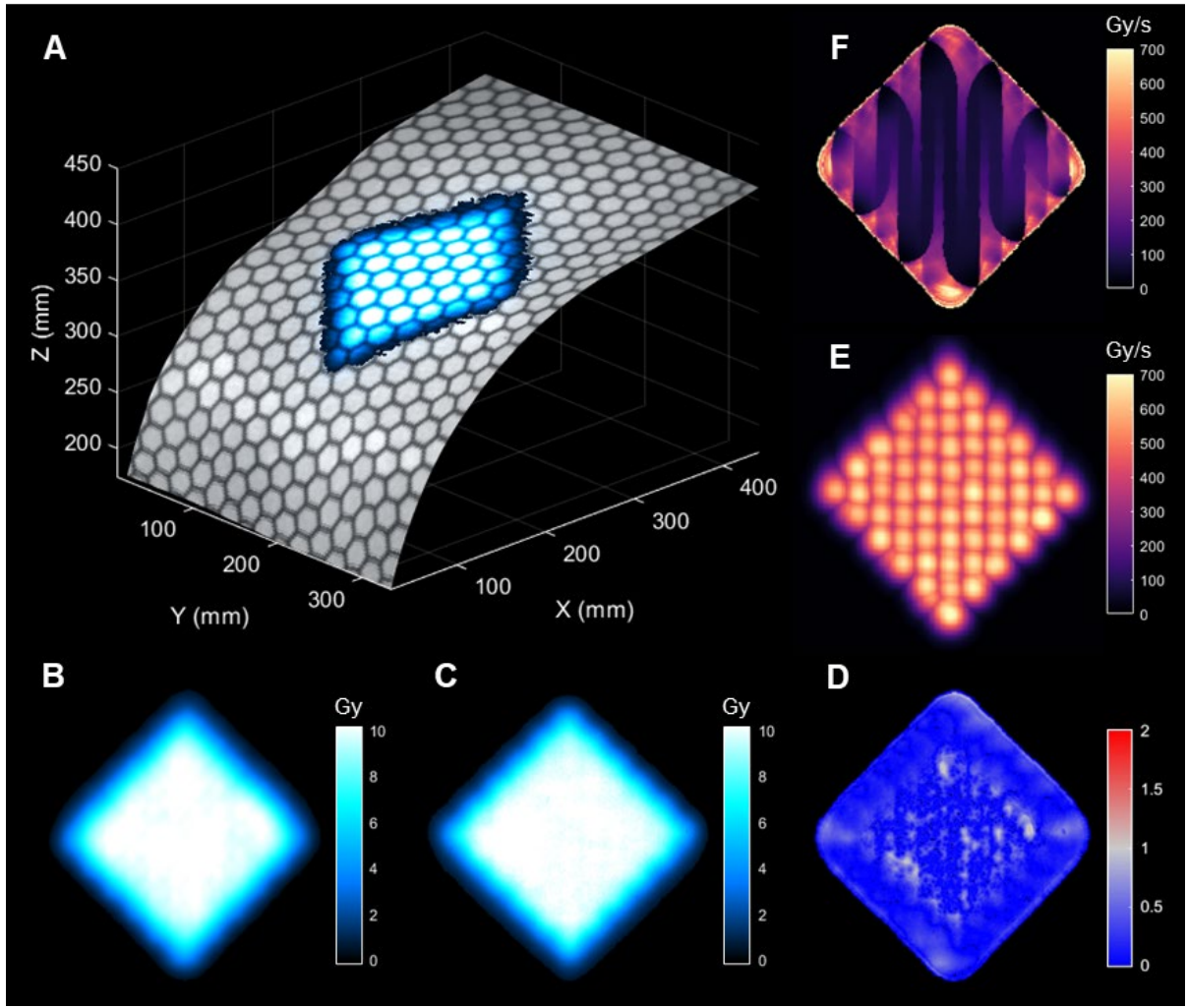

**Figure 9:** The resulting dataset of the scintillator array imaging technique is shown with an (a) overlay of cumulative scintillation profile over the surface mesh, (b) the corresponding corrected cumulative dose map, (c) a film delivery of the plan over a flat surface, and (d) 3%/2mm gamma comparison between the two. The (e) maximum and (f) PBS dose rate maps were also evaluated.

## 4. Discussion:

We developed a novel in vivo dosimetry system and workflow for monitoring full-field surface dose delivery in real time with a direct reference to patient surface anatomy. The key feature of the system is its ability to compare dose in the beam's-eye view, where stereovision cross-calibrated with a high-speed scintillation imaging camera makes it possible to evaluate the delivered profiles against the treatment plan, achieving <1mm accuracy at 1000 Hz and 99.8% duty cycle, with a maximum frame rate of 12 kHz[13]. Dose reporting is achieved using a modular scintillator array composed of rigid single emitter elements with a flexible connecting structure. This design facilitates adaptability without compromising spatial accuracy, which is crucial for in vivo applications. Emphasis in design was placed on ensuring spatial homogeneity across the scintillator elements and scalability of the array, making it easily adaptable to complex clinical translation.

Real-time 3D surface mapping and alignment of dose profiles to BEV was achieved via dual-module stereovision imaging where individual scintillator elements were identified in 3D space and used as reference points for downstream analysis. Following pre-treatment extrinsic and intrinsic calibration, the stereovision system accuracy was evaluated using a custom 52-point 3D phantom. Our stereovision-based QA procedure yielded an average localization error of 0.62 mm, satisfying the 1 mm threshold typically used in clinical practice. The novel 3D QA phantom further confirmed minimal deviations across the x, y, and z axes thereby supporting reliable geometric corrections and the capability to detect and address systematic offsets. Furthermore, the water-equivalent depth of the scintillator array was measured via dual energy build up testing to be 1.1 mm, which is minimal and suggests that the array can be included in treatment planning systems or simulations as a virtual bolus without major effects on dose distribution.

The system demonstrated high temporal resolution (~1 ms), enabling real-time tracking of dynamic PBS beam delivery, thereby addressing the challenge of capturing rapid changes in dose and dose rate during UHDR treatments. Dose reporting showed deviations within acceptable clinical trial standards when compared to film delivery, which supports the system's use for in vivo dose validation. Importantly, we developed robust corrections for angular dependencies of scintillator emission, ensuring accurate mapping of intensity to delivered dose. By accounting for variations in emission intensity due to viewing and irradiation angles, the developed angular corrections preserved dose accuracy across the treatment field, yielding a readout accuracy of ~1% (5.428 mGy). This is not feasible with any other proposed 1D or 2D dosimetry system since they either lack the ability to identify single element angles or are continuous which prevents discretized correction. Overall, a gamma analysis performed with 3%/2 mm criteria showed a 99.9% passing rate, presenting high agreement between measured and planned dose profiles, thereby demonstrating a robust validation suitable for clinical practice.

While the current system shows promising performance, future research will focus on expanding its application to more complex clinical plans involving spatially and temporally variable dose distributions, which would provide a more comprehensive validation of the system capability in a clinical setting. Additionally, the application of the array will be expanded to conventional treatments and other modalities such as photon and electron therapy. The impact of the array on photon and electron dose build-up and intensity variation with angular orientation will likewise be investigated. Potential applications include surface dose verification during breast cancer treatment to monitor contralateral dose delivery and head and neck treatments to quantify surface dose profile development throughout multi-fractionated treatments. Lastly, future efforts should investigate the integration of this system with existing quality assurance protocols and well as explore additional capabilities, such as monitoring local linear energy transfer, which may further facilitate making the proposed technique more broadly applicable in clinical use.

## 5. Conclusion:

We demonstrated a novel approach to in vivo beam monitoring using a freely deformable multi-element scintillator array coupled with passive surface imaging with linear dose-intensity response in a UHDR environment. By integrating high spatiotemporal resolution imaging with real-time stereovision geometric reconstruction, we presented an adaptable and scalable technique for surface dosimetry on geometrically complex surfaces. This work demonstrates its ability to accurately capture dose maps with high agreement between planned and imaged profiles, and calculate dose rate distributions. The methodology proves valuable for in vivo beam monitoring in proton therapy, offering real-time feedback and showing potential to enhance treatment accuracy and safety. Future research will focus on optimizing the system's performance and exploring applicability of this approach to other modalities.

## 6. Acknowledgements

This work was funded by NIH research grants R44CA268466 and P30CA023108. This work was also supported by grant #ROI2023-003 from the Radiation Oncology Institute.

## 8. Figure Captions

**Figure 1:** (a) Experimental setup showing the novel scintillator array over the patient treatment area, proton gantry delivering a PBS treatment, and an imaging system located at 2 m distance from target. Photo of the imaging system (b) consisting of high frame rate camera and two neighboring stereovision modules. Detailed photo of the scintillator array used for this study is shown (c) along with a breakdown of the element composition (d).

**Figure 2:** Image correction process is shown, using the (a) binarized profiles of the stereovision images to define a (b) 3D centroid point cloud. By translating the point cloud orientation to BEV, the centroid

markers are used as anchors to (c) deformably register the scintillation image into BEV via local weighted mean registration.

**Figure 3:** Method for angular correction is shown and applied to the (a) geometrically corrected scintillation image. (b) Using 3D information from stereovision surface difference between normal, gantry, and camera angles are determined and related to an intensity correction factor. The process is repeated to all elements containing scintillation signal to create a (c) correction factor image. Data for an arbitrary scintillating element 29 is shown as an example.

**Figure 4:** (a) Cumulative scintillation image post geometric and angular correction is shown undergoing single spot intensity interpolation. (b) An overlay of the kernel fit over the raw data for an arbitrary scintillation spot 219 is shown, with its location highlighted in the (c) fully interpolated scintillation image.

**Figure 5:** Histogram showing the cumulative distribution of mean-of-residual values for every spot in the 5x5 delivery. A t-location-scale function was used to fit the data (FWHM of 5.428 mGy), highlighting the enhanced tails due to overfitting of partial imaging frames and underfitting of motion frames.

**Figure 6:** Geometric correction procedure showing the system verification using the (a) QA stereovision phantom, (b) the corresponding image from a single stereo camera and (c) point cloud comparison of ground truth phantom locations to the stereovision reconstruction. Centroid localization error was evaluated showing the (d) Euclidean distance error histogram and the (e) single axis distributions.

**Figure 7:** Variation in scintillation intensity based on (a) observation angle and (b) irradiation angle was characterized and used to derive angular correction factors for the angular correction process outlined in section 2.7. The dependence of scintillation output on (c) dose was characterized and fit to a linear function showing high accuracy ($R^2 = 0.9983$).

**Figure 8:** Modelled setup used to characterize the water equivalent depth (WED) of the scintillator array by delivering beams of known energy to a multilayer ionization chamber detector. Tests at 160 MeV and 200 MeV similarly indicate a WED of 1.1mm.

**Figure 9:** The resulting dataset of the scintillator array imaging technique is shown with an (a) overlay of cumulative scintillation profile over the surface mesh, (b) the corresponding corrected cumulative dose map, (c) a film delivery of the plan over a flat surface, and (d) 3%/2mm gamma comparison between the two. The (e) maximum and (f) PBS dose rate maps were also evaluated.